# Shock generated vorticity in spark discharges

Bhavini Singh[1], Lalit K. Rajendran[1], Pavlos P. Vlachos[2] and Sally P.M. Bane[1*]

[1] Purdue University, School of Aeronautics and Astronautics, West Lafayette, USA.

[2] Purdue University, School of Mechanical Engineering, West Lafayette, USA.

*sbane@purdue.edu

**Abstract**

Spark discharges induce a complex flow field consisting of a shock wave at early times (~ 1 $\mu$s), a pair of vortex rings, and a hot gas kernel. The vortex rings entrain ambient gas into the hot gas kernel and control its cooling and expansion. In this work, we investigate the shock wave's contribution in producing the vortex ring vorticity. We analyze high-speed (700 kHz) schlieren images of the shock wave for a range of electrical energies to measure the shock properties and estimate shock velocity and curvature. These measurements are combined with a model to calculate the vorticity generated. The measurements show that the highest vorticity is generated near the peak shock curvature location, and the shock curvature and strength increase with electrical energy deposited. A comparison of the vorticity estimated from the model to vorticity measurements from stereoscopic particle image velocimetry shows the results to be statistically equivalent. This suggests that the shock curvature and velocity contribute to the vortex rings induced by spark discharges.

## 1. Introduction

Spark plasma discharges lead to rapid heating and pressure rise in the electrode gap, resulting in a shock wave [1]–[4]. The shock wave moves radially out of the center of the gap within a few microseconds [3], [5], [6] and decays rapidly into an acoustic wave. After the shock wave departure, the flow field is comprised of a hot gas kernel and vortex ring(s). The number of vortex rings, their dynamics, and the hot gas kernel evolution depend on the spark generation method [7]–[13]. In laser sparks, for instance, a pair of vortex rings of unequal strength and size are induced [8], [14], while in sparks generated between cone-tipped electrodes (pin-to-pin discharges), a pair of two almost identical vortex rings are induced [9]–[11] and in surface discharges, a single vortex ring is formed that propagates away from the discharge surface [12].

In these cases, the strength and dynamics of the vortex ring(s) control the spatial extent and cooling of the hot gas kernel and thus play a critical role in engineering applications. In laser sparks, which are typically used for combustion applications, the vortex ring characteristics control the direction and ejection of the hot gas kernel [8], affecting ignition and flame growth [15]. In pin-to-pin electrical spark discharges, the rings entrain cold ambient gas into the hot kernel to promote rapid cooling/mixing, and this determines the required pulsation frequency of the discharges to utilize the synergetic effect of multiple pulses to optimize ignition systems [16] or to stabilize flames and promote efficient combustion [17]–[19].

Despite the critical role vortex rings play in the flow induced by spark discharges, to the best of our knowledge, there has been no quantitative experimental investigation on the source of this vorticity. Researchers have used CFD simulations to study the flow field at early times during the spark induced shock wave, and have suggested that there are two possible sources of vorticity generation, both through a baroclinic effect: (1) vorticity production due to the non-uniformity in the shock strength [3], [20], [21] (2) vorticity production due to the interaction of pressure gradients behind the shock wave with the hot gas kernel [7], [8], [20], [22]–[24]. In laser sparks, the second mechanism has been shown to be the dominant source of vorticity generation because the blast wave expands out of the field of view, depositing vorticity rapidly, while the pressure gradients and expansion waves following the blast wave interact with the low density region from the hot gas kernel for longer periods of time and generate more vorticity than the shock wave. In nanosecond spark discharges, the energy deposited in the gap is an order of magnitude lower than in laser sparks and the induced shock wave and trailing pressure gradients and expansion waves are also weaker. The two mechanisms by which vorticity is generated therefore may occur at similar time scales, and it is possible that because the trailing pressure waves are weaker in nanosecond sparks, the vorticity they generate is also weak. Therefore, the contribution of the vorticity jump across the shock should not be overlooked and a detailed experimental analysis on the contribution of both mechanisms in low energy plasmas is needed. In this work, we experimentally investigate the shock wave's contribution to the vortex ring vorticity generation, via the first mechanism, as an initial estimate of the vorticity. We employ high-speed schlieren visualization and image processing methods to measure the shock properties and a model to calculate the vorticity based on these shock properties. We compare the estimated vorticity to measurements from stereoscopic particle image velocimetry (S-PIV) from prior work [11] and explore how the shock wave accounts for the vorticity in the vortex rings that drive cooling of the hot gas kernel at later times.

## 2. Vorticity generation due to a curved shock

Vorticity can be generated at a boundary and within a fluid due to baroclinicity. The vorticity transport equation is given as:

$$\frac{D\vec{\omega}}{Dt} = (\vec{\omega} \cdot \nabla)\vec{u} - \vec{\omega}(\nabla \cdot \vec{u}) + \frac{1}{\rho^2}\nabla\rho \times \nabla p + \nabla \times \left(\frac{1}{\rho}\nabla \cdot \tau\right) \quad (1)$$

where $\vec{\omega}$ is vorticity, $\vec{u}$ is the velocity, $\rho$ and $p$ are the density and pressure, respectively and $\tau$ is the viscous stress tensor. The first two terms on the right-hand side represent changes in vorticity due to vortex stretching and compressibility effects, respectively. The third term represents the generation of vorticity due to misalignment of pressure and density gradients, which leads to baroclinic torque. The last two terms represent the diffusion of vorticity due to viscosity and the generation of vorticity due to a body force. In a compressible flow, such as the early stages (< 20 $\mu$s) of the flow induced by a spark discharge, if we use the simplifying assumption that the flow is inviscid, we can neglect the last term in Equation 1.

In spark discharges, one source of vorticity is baroclinic torque arising from the non-uniform strength of the shock wave [7], [8], [20], which results in a curved shock. The generation mechanism of vorticity behind a curved shock has been explained using a purely dynamic approach by Hayes[25], who shows that the vorticity generated due to a curved shock is dependent only on the magnitude of the tangential component of the shock velocity. Other work used Crocco's theorem to arrive at the same result as Hayes for vorticity jump across a curved shock [26]–[28]. These works show that the non-uniformity in shock strength generates non-uniform entropy behind a curved shock, resulting in an entropy gradient along the shock, which is responsible for vorticity generation. However, these studies are restricted to steady flows due to Crocco's vorticity law's inherent assumption, a restriction that is not necessary for Hayes' derivation.

The schematic in Figure 1 shows a slice of the curved shock generated by a spark discharge in cylindrical coordinates. The middle portion of the shock is approximately cylindrical and has been shown, via CFD simulations, to be stronger than the portions of the shock closer to the electrodes [3]. The difference in the strength of the shock along its length results in pressure and density gradients behind the shock that are misaligned, leading to baroclinicity [29]–[33]. The misalignment in the density and pressure gradients occurs in the regions where the shock is curved, with almost no misalignment in the center of the gap where the shock wave is mostly straight.

According to Hayes [25], the vorticity jump across this unsteady, axisymmetric, curved shock, generated under quiescent flow conditions, is given by:

$$\omega = -\frac{\partial U_s}{\partial e_1}(\epsilon - 1)^2 \epsilon^{-1} \tag{3}$$

where $U_s$ is the normal shock velocity. The tangential gradient $\frac{\partial U_s}{\partial e_1}$ can be expressed in terms of the tangential shock velocity, $U \cos\sigma$, and shock curvature, $\kappa$, hence Equation (3) becomes:

$$\omega = -U \cos\sigma \, \kappa (\epsilon - 1)^2 \epsilon^{-1} \tag{4}$$

where $\sigma$ is the angle the shock makes with the horizontal. The local shock curvature ($\kappa = d\sigma/de_1$) is negative for the convex surface of the spark induced shock. The shock strength, measured as the density ratio $\epsilon$, is the ratio of the density behind the shock, $\rho_s$, to the density ahead of the shock $\rho_\infty$:

$$\epsilon = \frac{\rho_s}{\rho_\infty} = \frac{(\gamma + 1)M_n^2}{(\gamma - 1)M_n^2 + 2} \tag{5}$$

where $\gamma$ is the ratio of specific heats, which assuming a perfect gas with constant specific heats, is taken to be 1.4. The normal shock Mach number, $M_n$, is calculated as $U_s/a$ where $a$ is the speed of sound in the ambient air, taken to be 343 m/s. Therefore, to calculate the shock-generated vorticity, the model (Equations 4-5) requires the density ratio and the tangential shock velocity estimated from the shock Mach number, and the shock curvature and shock angle obtained from the shock profile. The next section details the experimental measurements capturing the shock wave and the procedure to estimate these parameters.

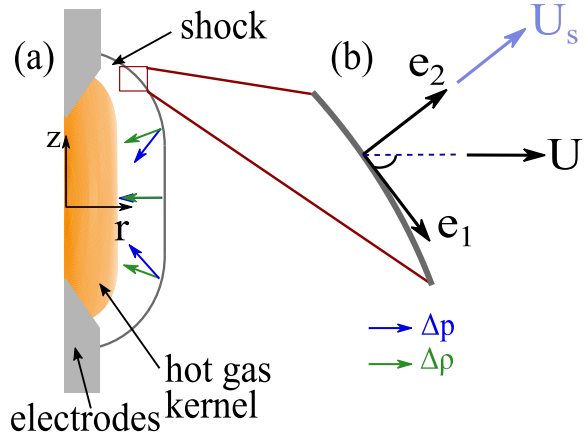

Figure 1: Schematic of shock generated due to a spark discharge showing (a) the curved nature of the shock wave in cylindrical coordinates and the misalignment of pressure and density behind the shock due to non-uniform shock strength; (b) the local coordinate system for a segment of the shock with normal shock velocity $U_s$ and horizontal component of shock velocity $U$.

## 3. Experimental methods and techniques

The shock wave induced by a single spark discharge was captured using schlieren imaging for energy values ranging from 4.3 to 5.3 mJ, resulting in a total of 10 distinct experiments. Each spark discharge was generated using a nanosecond high voltage pulser in a 5 mm gap between two ceriated-tungsten electrodes with cone-shaped tips. Details on the plasma generation are given in prior work [10]. The experiments were conducted at least 30 seconds apart to eliminate residual flow effects from one spark event to the next. The parameters needed to calculate the vorticity jump were extracted from the first image of the shock after it had separated from the hot gas kernel, approximately 3.5 $\mu$s after the spark.

### 3.1 Schlieren Imaging

A schematic of the Z-type schlieren set-up used to image the spark induced shock wave is shown in Figure 2. The system consisted of a 150 W xenon arc lamp (Newport 66907) and a 60 mm aspheric condenser lens to create a point light source. Two concave mirrors 152.4 mm in diameter and with a focal length of 3.05 m, were used to collimate the light beam then converge it onto a knife edge. An intermediate lens was used to reduce light spillage before the collimating mirror to control the ray cone's angle emitted from the light source. The camera was set to a frame rate of 700,000 fps (~ 1.4 $\mu$s between frames) at a resolution of 64 x 128 pixels corresponding to a field of view of approximately 5 x 10 mm and magnification of 0.08 mm/pixel. A delay generator was used to synchronize the camera and the high voltage pulser.

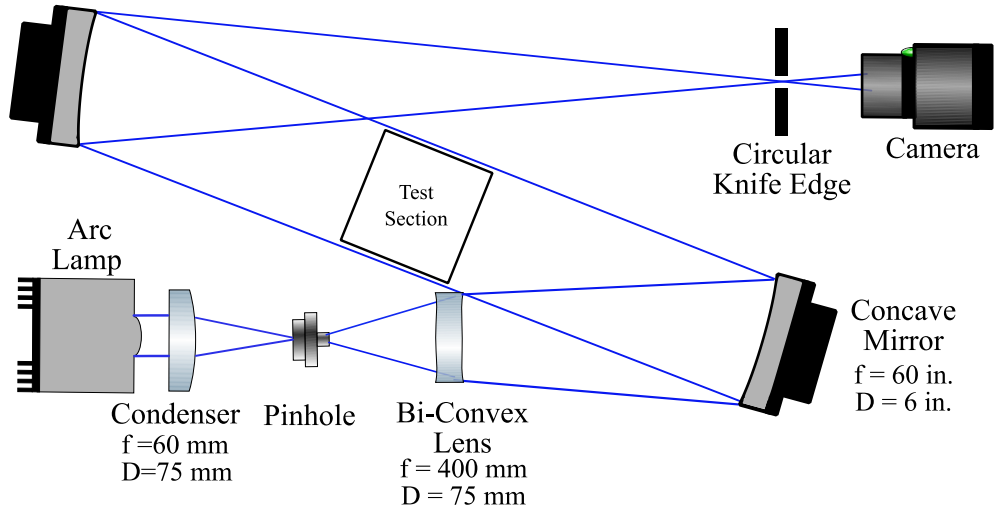

Figure 2: Schematic of the experimental set-up for schlieren measurements of the plasma-induced flow field

### *3.2    Processing metrics and calculation of parameters*

A correlation-based image processing method was used to estimate the shock parameters. The ultra-high-speed recording leads to high image noise due to low exposure times and low spatial resolution due to camera hardware limitations. These two challenges lead to competing requirements in the estimation of the model parameters that involve derivative calculations. These derivative calculations can amplify the already high noise and deteriorate the measurements due to large random errors. However, reducing the noise by simple smoothing methods is also challenging because they reduce the measurements' spatial resolution and increase systematic errors in the derivative estimates. Therefore, a data processing procedure was employed to achieve an optimal tradeoff between the measurements' systematic and random errors.

The processing steps used to model the vorticity generated due to shock curvature from the schlieren images are summarized in Figure 3. Each schlieren experiment captured 990 pre-triggered images of the background and 10 images after the spark discharge. The mean of the pre-triggered images was subtracted from the first image of the shock after it had separated from the hot gas kernel, approximately 3.5 $\mu$s after the spark. The background-subtracted image was then flipped along the center of the electrode gap and averaged. The new shock image was used to detect the shock outline and calculate the shock velocity.

### *3.2.1   Shock outline detection*

The shock outline was detected using a Gaussian subpixel fitting scheme. Since these methods are sensitive to noise, we developed a cross-correlation based method to mitigate this effect, as image noise is expected to be uncorrelated across pixels in the image. To detect the shock outline, each row of pixels in the shock image was cross-correlated with the first row of pixels containing the shock by first replicating the single row of pixels to create a 2D intensity map which was 16 pixels in height (Figure 3 (a)) . We applied a 50% Gaussian window [34] to the 2D map of intensities and cross-correlated the intensities using Robust Phase correlation [35], [36] to determine the displacement of the shock and define the shock outline. The ensemble cross-correlation was calculated along the columns of each 2D map, and a least-squares Gaussian fit to the cross-correlation peak was used to obtain the subpixel location of the shock profile along the horizontal direction for each row of pixels. We then smoothed the shock profile using a moving linear regression.

We used the shock profile (Figure 3(b)) to calculate the shock angle and shock curvature. The shock angle (Figure 3(c)) was calculated as the arctangent of the derivatives of the $r$ and $z$ locations of the shock outline, with the derivatives calculated using a 4$^{th}$ order compact noise optimized Richardson extrapolation scheme[37]. The shock angle and the Euclidean distance between velocity grid points were used to calculate the shock curvature (Figure 3(d)).

### *3.2.2 Shock velocity calculation*

The shock velocity was calculated similarly to the shock outline, using cross-correlation based methods to mitigate the effect of image noise and Gaussian subpixel fitting schemes to calculate the subpixel shock displacement. To determine the normal shock velocity, we cross-correlated successive shock images recorded at 3.5 $\mu$s and 4.9 $\mu$s after the spark. The interrogation windows had a height of 16 pixels, containing 16 points of the shock outline and the orientation of the window was determined using the slope of the line perpendicular to the center shock point in the window. We applied a 75% overlap and a 50% Gaussian window to the original window size, resulting in a window resolution of 8 pixels in height and grid resolution of 2 pixels corresponding to 0.16 mm. The intensities in the same window from the first and second time steps were cross-correlated using Robust Phase correlation to determine the normal shock velocity. The ensemble of the cross-correlation plane obtained for the shock velocity calculation was calculated, and a least-squares Gaussian fit to the cross-correlation peak was used to determine the subpixel shock displacement.

### *3.2.3 Shock-generated vorticity calculation*

The tangential shock velocity (Figure 3(f)), shock angle, shock curvature, and the density ratio (Figure 3(g)), which was calculated from the normal shock Mach number, were used to calculate the spatial variation of the vorticity jump across the shock (Figure 3(h)) according to Equation 4. All results were non-dimensionalized using the induced velocity and density behind the shock, calculated according to [9], [38], and the characteristic length scale was taken to be half the electrode gap distance (2.5 mm). Further details on the non-dimensionalization can be found in prior work [10].

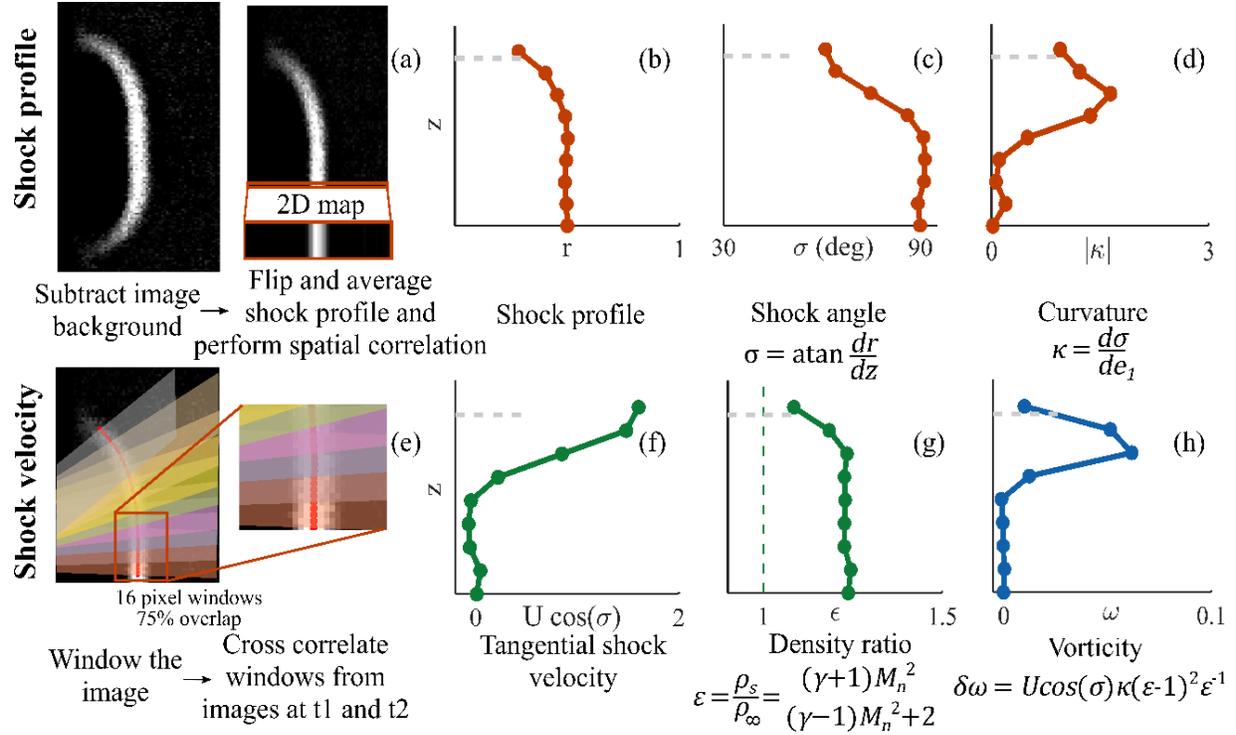

Figure 3: Processing steps used to calculate the vorticity jump across a shock wave using images obtained from schlieren experiments. The z-location of the electrode tip is shown as a gray dashed line.

### 3.3 Uncertainty quantification

As mentioned earlier, the high image noise levels and derivative calculations lead to an amplification of the uncertainties in the vorticity estimates. Experimental uncertainties in the displacement field used to estimate both the shock profile and the shock velocity were calculated using the moment of correlation (MC) method [39], then propagated through the shock-generated vorticity calclations using the Taylor series-based propagation method[40]. The maximum uncertainties in the calculation of $r$ and $z$ for the shock profile were 0.03 mm and 0.04 mm, respectively, corresponding to 1.2 % and 1.6 % of the half-gap distance (2.5 mm). The maximum uncertainties for the shock velocity calculation were 20 m/s, corresponding to 5 % of the maximum velocity, 412 m/s. The modeled vorticity jump uncertainties ranged from 20% to 60% of the vorticity jump value, with the majority (~70%) of the vorticity jump uncertainties being less than 40% of their respective vorticity jump values.

## 4. Results and discussion

### 4.1 Shock wave induced by spark discharge

An example of the flow field induced after a single spark discharge generated under quiescent conditions in a 5 mm electrode gap and corresponding to a deposited energy of 4.8 mJ is presented in Figure 4(a). The flow field shown spans approximately 10 $\mu$s and is comprised of a cylindrical hot gas kernel and a shock wave, trailed by expansion waves. In the first snapshot at $t/\tau = 0.3$, the shock wave has completely separated from the induced hot gas kernel. The shape of the shock wave resembles that shown in the schematic in Figure 1, where the parts of the shock closest to the electrodes are curved, while the parts near the center of the electrode gap are cylindrical. The normal Mach number along the shock ranges between 1.05 to 1.2, with the normal shock velocity decreasing along the length of the shock from the center of the electrode gap out toward the electrode tips. The Mach number in the center region of the gap remains approximately constant at 1.2. The shape of the Mach number profile (not shown) follows that of the shock outline and resembles the density jump profile shown in Figure 5(d).

The first time step shown in Figure 4(a) is used to calculate the vorticity jump across the shock because the progressive reduction in the magnitude and non-uniformity of the shock velocity will lead to negligible vorticity production at later times. The shock-generated vorticity estimated using the model is compared to the vorticity calculated directly from a separate set of S-PIV experiments from prior work [11]. An example of the vorticity field induced at later times is shown in Figure 4(b), where a pair of vortex rings are generated near the electrode tips.

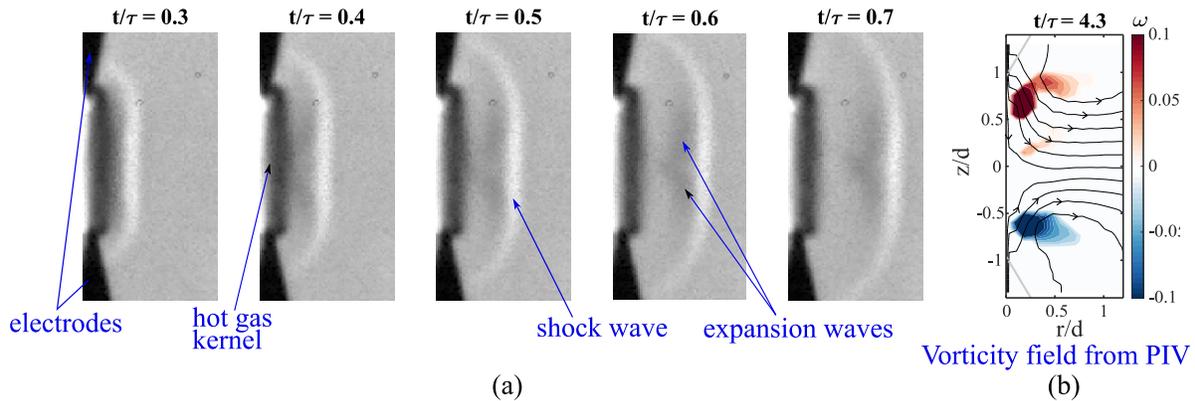

Figure 4: (a) Time evolution of a shock wave induced by a single spark discharge (4.9 mJ energy and $\tau = 14\ \mu s$) between two electrodes (5mm gap). (b) Vorticity field measured in prior work for a similar spark discharge (5 mm gap, 4.8 mJ, $\tau = 14.1\ \mu s$) showing a pair of vortex rings generated near the electrode tips[11].

### *4.2 Vorticity estimation based on shock curvature*

The spatial profiles of the parameters used to calculate the vorticity jump across the spark induced shock are shown in Figure 5 and normalized by their maxima. The profiles calculated from each schlieren experiment are overlaid in the same figure. The orange profiles show quantities calculated using the shock profile, the green profiles show quantities calculated using a combination of the shock profile and shock velocity, and the blue profiles show the vorticity.

The shock outlines in Figure 5(a) are seen to be curved in all tests, with the curvature in Figure 5(b) being close to zero near the center of the gap and highest near the electrodes (the z-location of the electrode tip is shown by the gray dashed line). The tangential shock velocity (Figure 5(c)) increases monotonically from the center of the shock ($z = 0$) to the ends of the shock near the electrodes. The density ratio term (Figure 5(d)), which represents the shock's strength, is almost constant near the center region of the shock and decreases as we move closer to the electrodes, corresponding to a decrease in the normal Mach number. The density ratio profile is the same as that of the normal shock velocity (not shown), both roughly resembling the shock profile. Combining the first 4 terms into Equation 4 gives the final vorticity jump across the length of the shock (Figure 5(e)), with negligible vorticity generated near the center of the gap and the peak vorticity being generated in the region of highest curvature.

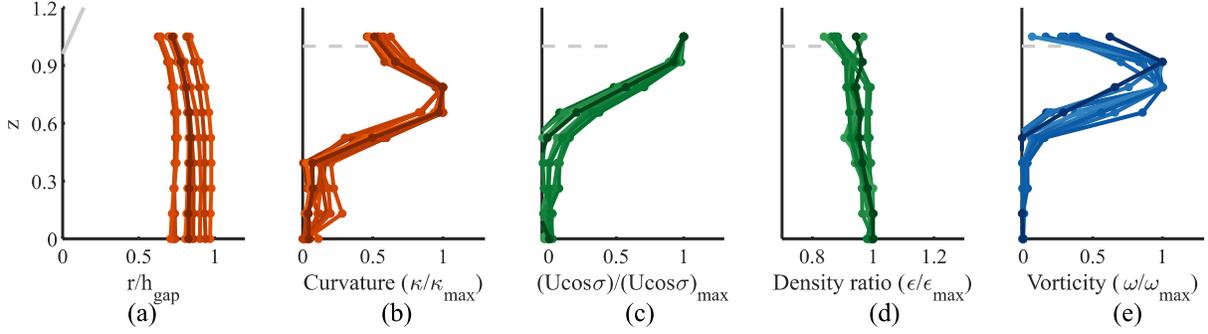

Figure 5: Normalized spatial variation of shock properties: (a) shock profile, (b) curvature, (c) tangential velocity, (d) density ratio, and (e) vorticity jump across the shock. Orange profiles are calculated using the shock profile, green profiles using a combination of the shock profile and the shock velocity. The z-location of the electrode tip is shown as a gray dashed line.

*4.3  Comparison of shock-estimated vorticity to experimental results from S-PIV*

For all the experiments, the peak vorticity is determined for comparison to the S-PIV results from prior work. The shock parameters calculated from the schlieren and S-PIV measurements are plotted against the analytic, normalized pressure gradient across the shock, calculated from the energy deposited in the electrode gap. This allows for comparing two independent measurements while demonstrating the effect of the analytic shock strength and energy deposited in the gap on the different parameters. The analytic, normalized pressure gradient, like the density ratio calculated in the previous section, is also a measure of the strength of the shock wave, although the density ratios calculated in this study are all obtained directly from the schlieren images. The analytic, normalized pressure gradient and analytic shock strength will be used interchangeably.

The analytic, normalized pressure gradient across the shock is related to the electrical energy deposited in the electrode gap and is calculated according to Jones et al. [38] as:

$$\frac{p_2 - p_1}{p_1} = \frac{2\gamma}{\gamma + 1}(M_1^2 - 1) = \frac{2\gamma}{\gamma + 1} \frac{0.4503}{\left(1 + 4.803 \left(\frac{r}{R_0}\right)^2\right)^{\frac{3}{8}} - 1} \tag{6}$$

where the subscripts 1 and 2 represent conditions upstream and downstream of the shock, respectively, and $r$ is the radial distance at which flow properties behind the shock are measured. In the present context, the radial distance is half the electrode gap distance (2.5 mm). The variable $R_0$ is the characteristic radius determined by the initial conditions and is given by:

$$R_0 = \left[\frac{4E_0}{3.94\gamma p_1}\right]^{\frac{1}{2}} \tag{7}$$

where $E_0$ is the electrical energy deposited by the plasma per unit length of the electrode gap. The analytic, normalized pressure gradient (analytic shock strength) increases with an increase in electrical energy deposited in the gap.

Figure 6 shows the peak vorticity and peak curvature locations and the shock curvature, tangential velocity, and density ratio which determine the peak vorticity as a function of the analytic pressure gradient. The peak vorticity location and the tangential shock velocity at peak vorticity show no clear dependence on the analytic shock strength. The peak vorticity location is roughly the same for almost all experiments, occurring near the peak curvature location. The shock curvature at the locations of peak vorticity show a weak dependence on the analytic shock strength, while the density ratio increases with the analytic shock strength, as expected.

These results show that the density ratio estimates calculated from the schlieren images can capture the expected increase in density ratio, which is a measure of the experimentally calculated shock strength, with an increase in the electrical energy deposited. We also demonstrate using these results that stronger shocks tend to have higher maximum curvature (smaller radius of curvature) than weaker shocks. The curvature values in the region of peak vorticity range from approximately 1 to approximately 1.7.

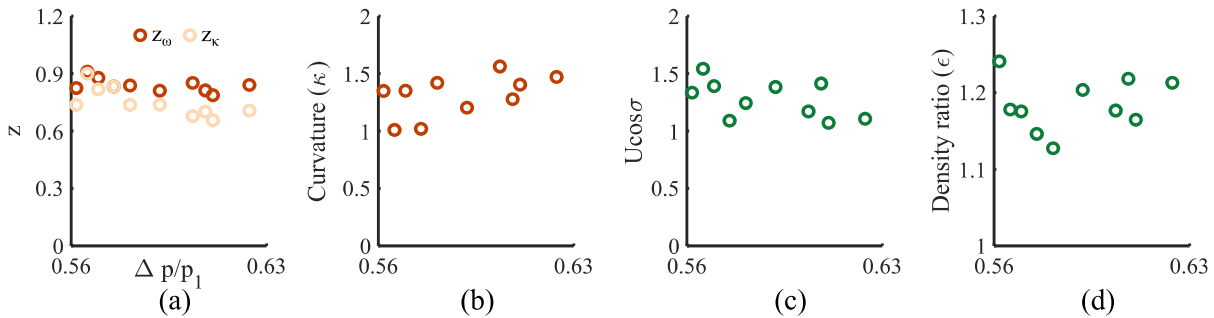

Figure 6: (a) The locations of peak vorticity ($z_\omega$) and peak shock curvature ($z_\kappa$) and (b) shock curvature, (c) tangential velocity, and (d) density ratio values used to calculate the maximum vorticity jump across the shock as a function of the analytic, normalized pressure gradient.

The peak shock-generated vorticity jump obtained from the schlieren images is compared to the mean vorticity within the vortex rings from S-PIV experiments described in prior work [11]. The velocity field obtained from S-PIV was flipped and averaged along the $z = 0$ axis to obtain a single coherent structure representing the vortex ring's core in each experiment. The peak shock-generated vorticity and the mean vorticity calculated in the vortex rings are compared in Figure 7. The median values are 0.046 and 0.05 from the shock and vortex rings, respectively. This shows that the vorticity from the model based on shock curvature and the vorticity in the vortex rings are statistically equivalent over the range of experiments performed in this study. This suggests that the shock curvature contributes to the vorticity observed in the vortex rings at later times.

However, there is considerable scatter in the vorticity values and a negligible dependence on the analytic pressure gradient. Further, the S-PIV and schlieren measurements are from different experiments, and some of the variability is likely due to the stochastic nature of the spark discharge as well as the high uncertainties in the shock-generated vorticity estimates. Additionally, the schlieren measurements are taken at approximately 3 µs, whereas the first PIV measurements are taken at approximately 30 µs and local values of vorticity can be altered post generation [41], further explaining the model and experiment's mismatch and limiting a direct comparison.

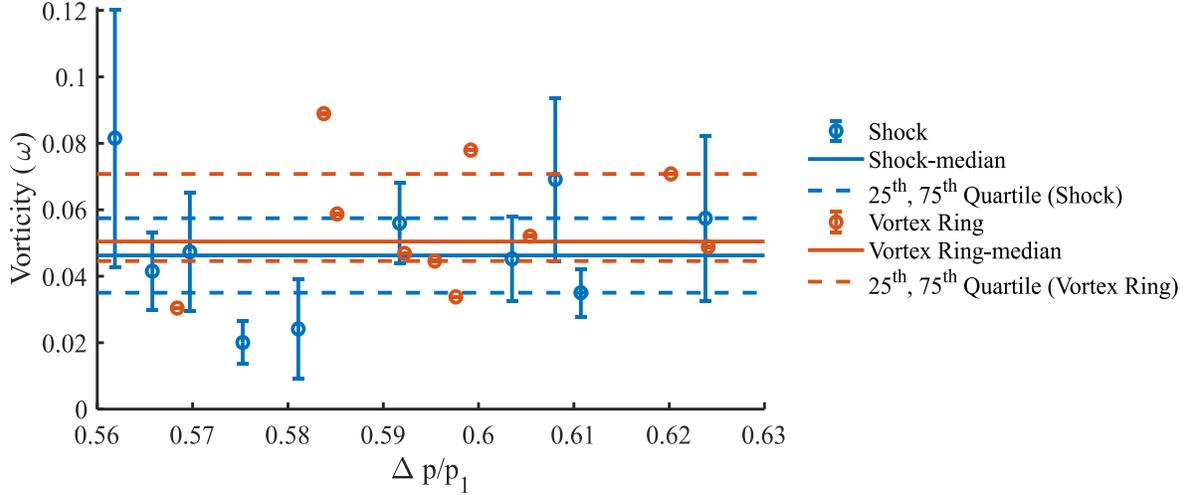

Figure 7: Comparison between the non-dimensionalized peak vorticity estimated from the schlieren images with a curved shock model and mean vorticity from PIV measurements.

## 5. Conclusions

The flow field induced by a single spark discharge consists of a curved shock wave that propagates radially outward, leaving behind a region of hot gas and vortex ring(s) which drive the cooling and expansion of the hot gas kernel. We use schlieren experiments to capture the shock and develop an image-processing procedure that mitigates the effects of high image noise and low spatial resolution to extract the shock properties. We use the shock properties to estimate the shock-generated vorticity and compare it to the vortex ring vorticity.

The ultra-high-speed recording (700,000 fps) used to obtain the schlieren images leads to high image noise levels, and hardware limitations lead to the low spatial resolution of the recorded images. To achieve an optimal tradeoff between the systematic and random errors in the measurements, we employ a cross-correlation based image processing procedure to extract the shock outline and shock velocity and use Gaussian subpixel fitting schemes commonly used in PIV to provide high spatial resolution. The shock outline is used to calculate the shock curvature and shock angle, while the shock velocity is used to calculate the density ratio, all of which are inputs to the model used to estimate the vorticity jump across the shock wave.

This work shows that the shock profile is approximately cylindrical in the center of the gap with nearly zero curvature, with the shock curvature increasing towards the electrodes and reaching peak curvatures of approximately $\kappa/\kappa_{gap} = 1.7$. The normal shock Mach number is highest near the center of the gap, close to Mach 1.2, and decreases near the electrodes to approximately 1.05. The peak values of both the shock Mach number and curvature increase with the analytic pressure gradient, which is related to the electrical energy deposited in the gap. The magnitude of the tangential velocity increases along the shock, and the peak tangential velocity shows no clear dependence on the analytic pressure gradient. The vorticity calculated using the model is close to zero at the center of the gap and progressively increases along the shock towards the electrode tips. The highest vorticity is observed in the regions of the highest shock curvature, with the peak vorticity showing no clear dependence on the analytic pressure gradient. These results are similar to CFD simulations [20] showing the presence of vortices in the region where the shock wave curvature is highest.

The shock-generated vorticity is compared to the vorticity from experimental S-PIV results from previous work [11] with both shown to be statistically equivalent. This suggests that the shock wave induced at early times ($< 20$ $\mu$s) contributes to the vorticity in the vortex rings that control the cooling and dynamics of the hot gas kernel at later times ($> 50$ $\mu$s). The work is limited in that the schlieren and PIV measurements were non-simultaneous, and the time at which the vorticity is calculated from the model is an order of magnitude lower than the first snapshot of vorticity from the PIV experiments.

We show that the vorticity induced by a spark discharge may be approximated from the shock curvature and velocity using qualitative schlieren experiments and used to obtain an initial estimate of the strength of the vortex rings, having implications in the late stages of the flow field. In the case of pin-to-pin discharges, an estimate of the vortex rings' strength can be used to approximate the rate of cooling [10], [11]. Controlling the strength and location of peak curvature of the shock wave induced at early times can affect the late stages of flow field dynamics.

[41] S. Ghosh and K. Mahesh, "Numerical simulation of the fluid dynamic effects of laser energy deposition in air," *J. Fluid Mech.*, vol. 605, pp. 329–354, 2008.15